\documentclass[twocolumn,english,aps,prl,superscriptaddress]{revtex4-1}
\usepackage[latin9]{inputenc}
\usepackage{amsmath}
\usepackage{amssymb}
\usepackage{graphicx}
\usepackage{esint}

\usepackage{ulem}

\makeatletter

\usepackage{graphicx}
\usepackage{color}
\usepackage[usenames,dvipsnames]{xcolor}
\usepackage[colorlinks=true,linkcolor=Red,citecolor=Green,linktoc=page]{hyperref}
\usepackage{multirow}
\usepackage{float}
\usepackage{flushend}
\usepackage{balance}
\usepackage[varg]{txfonts}
\usepackage{ulem}
\usepackage{fancyhdr}

\makeatother

\usepackage{babel}
\begin{document}
\title{Reply to P. Ao's Comment on ''Sign reversing Hall effect in atomically thin high temperature superconductors''}
\author{S.\,Y.\,Frank\,Zhao}
\affiliation{Department of Physics, Harvard University, Cambridge, MA 02138, USA}
\author{Nicola\,Poccia}
\affiliation{Department of Physics, Harvard University, Cambridge, MA 02138, USA}
\author{Margaret\,G.\,Panetta}
\affiliation{Department of Physics, Harvard University, Cambridge, MA 02138, USA}
\author{Cyndia\,Yu}
\affiliation{Department of Physics, Harvard University, Cambridge, MA 02138, USA}
\author{Jedediah\,W.\,Johnson}
\affiliation{Department of Physics, Harvard University, Cambridge, MA 02138, USA}
\author{Hyobin\,Yoo}
\affiliation{Department of Physics, Harvard University, Cambridge, MA 02138, USA}
\author{Ruidan\,Zhong}
\affiliation{Department of Condensed Matter Physics and Materials Science, Brookhaven National Laboratory, Upton, New York 11973, USA}
\author{G.\,D.\,Gu}
\affiliation{Department of Condensed Matter Physics and Materials Science, Brookhaven National Laboratory,
    Upton, New York 11973, USA}
\author{Kenji\,Watanabe}
\affiliation{National Institute for Materials Science, Namiki 1-1, Tsukuba, Ibaraki 305-0044, Japan}
\author{Takashi\,Taniguchi}
\affiliation{National Institute for Materials Science, Namiki 1-1, Tsukuba, Ibaraki 305-0044, Japan}
\author{Svetlana\,V.\,Postolova}
\affiliation{Institute for Physics of Microstructures RAS, Nizhny
Novgorod 603950,Russia}
\affiliation{Rzhanov Institute of
Semiconductor Physics SB RAS, Novosibirsk 630090, Russia}
\author{Valerii\,M.\,Vinokur}
\affiliation{Materials Science Division, Argonne National Laboratory, Argonne, IL 60439, USA}
\affiliation{Consortium for Advanced Science and Engineering, Office of Research and National Laboratories, University of Chicago, Chicago, IL 60637, USA}
\author{Philip\,Kim}
\affiliation{Department of Physics, Harvard University,
    Cambridge, MA 02138, USA}

\maketitle


\begin{figure}[t!]


	\begin{center}
		\includegraphics[width=1\linewidth]{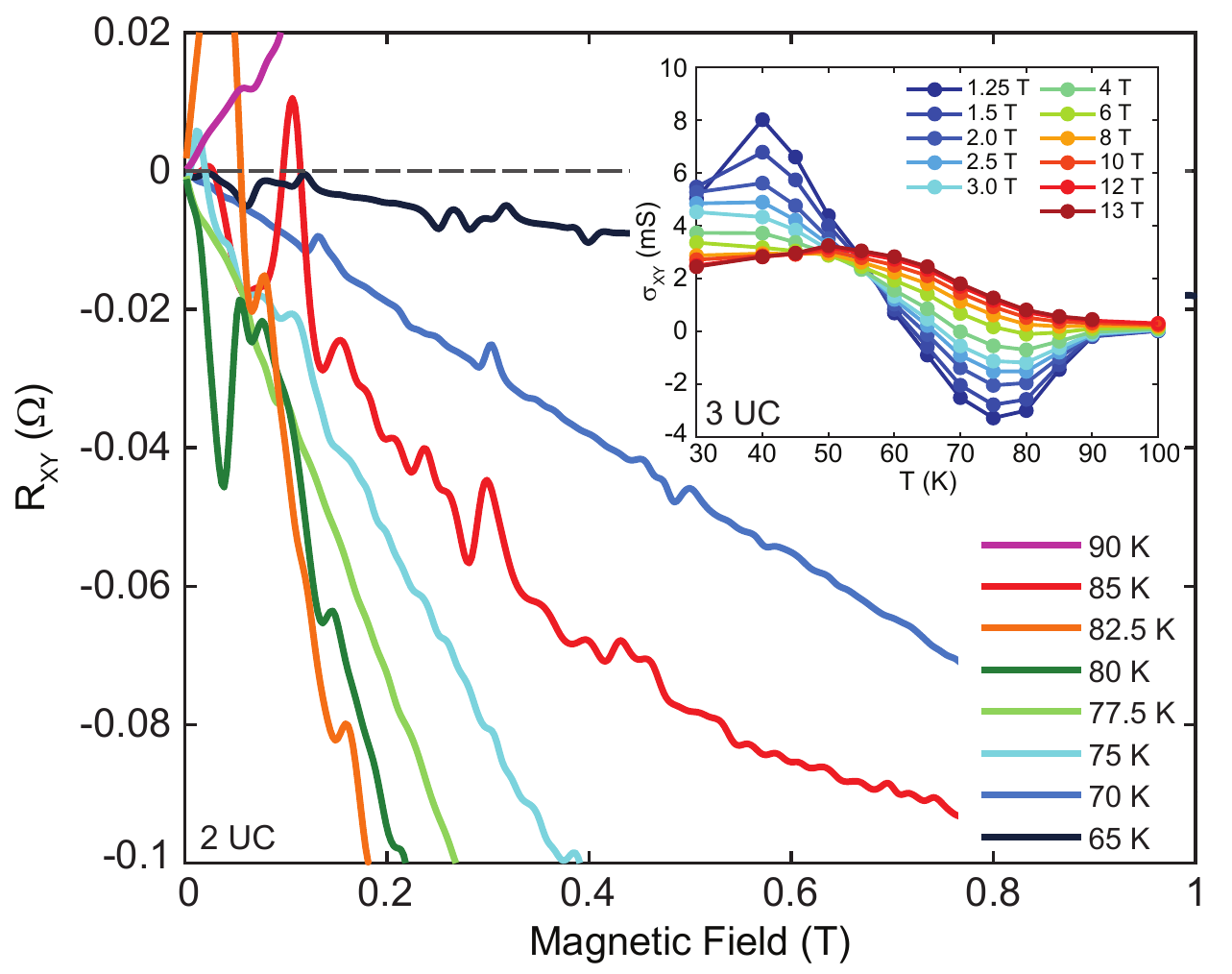}

		\setlength{\belowcaptionskip}{-20pt}

		\vspace{-0.4cm}		

		\caption{\textbf{Hall Sign Reversal} for a 2.0 unit cell thick Bi-2212 crystal at low magnetic fields. We observe no evidence of a lower critical field, below which the sign reversal vanishes. \textbf{Inset:} Temperature dependence of the Hall conductivity at various constant magnetic fields for a 3 UC device.}

		\label{Fig1}

	\end{center}

\end{figure}


The essence of Ao's theory, as he puts it in\,\cite{AoJPCM} is that ``...the Hall anomaly can be understood based on the vortex vacancy motion in a pinned vortex lattice." However, the temperature interval where we observe the sign reversal falls mostly into the vortex-liquid regime where vortex lattice, let alone vortex vacancies, do not exist. In simple words, Ao's theory is not related to the reality of the physical world. 

The inability of Ao's theory to address the experiment can be illustrated comparing two specific predictions made in Ao's comment to our observations:

\vspace{-0.3cm}
\begin{enumerate}

	\item According to\,\cite{AoC,AoJPCM,AoJS}, a ``lower critical field" should emerge, below which the Hall sign reversal should vanish. However, as seen in Figure 2a in our paper\,\cite{OurPaper} 
	which we re-plot here in Fig\ref{Fig1}, specifically for low magnetic fields,  the Hall resistance is negative (i.e. reverses in sign) at all nonzero magnetic fields, without any hint to the existence of the lower critical field. 
	 Neither was lower critical field reported in previous experiments.

	\item Likewise, our data do not support the second prediction of\,\cite{AoC} that the Hall conductivity follows an ''Arrhenius law'' with an activation energy corresponding to energy for the generation of ''vortex vacancies in the vortex lattice'' \cite{AoJPCM, AoC}. As seen in Figure\,\ref{Fig1} Inset, the Hall conductivity does not evolve monotonically with temperature, much less follows an Arrhenius law. In the Supplementary of\,\cite{OurPaper}, we show that the Hall sign reversal disappears in our samples below the BKT transition around 60 K. Thus sign reversal indeed exists only in the vortex liquid regime where vortex lattice and, therefore, vortex vacancies simply do not exist. Our data are in full agreement with the findings of all other experiments where the sign reversed Hall effect has been mostly seen in the vortex liquid regime, and where the Hall signal vanishes as the vortex liquid freezes into a solid.

\end{enumerate}

A detailed look at references in Ao's Comment \cite{AoC} refutes his claim that his theory was supported by data from various laboratories. Of the 13 papers Ao referenced (Refs. 6-18) in\,\cite{AoC}, \textbf{only one} (Ref.[11]) quantitatively compared Ao's findings to experiment and stated that the observed experimental behavior of Hall conductance \textbf{deviates from predictions specific to his theory}. Reference [8] indicated that Ao's theory is inapplicable to their experiment, and Refs.\,[9, 10, 12, 13, 15-18], simply mentioned Ao's work in passing as one of many citations but did not present any data to support Ao's theory. Only Ref.\,[14] claimed in a single sentence that Ao's theory explains their experiment but did not make any quantitative comparison. And, finally, Ref.\,[6], which according to Ao ''explicitly tested against data published in PRL'' \cite{AoC}, is Ao's one-page Comment not presenting a single equation or fit to the experiment.

\nocite{*}

\end{document}